# The structure and motion of incoherent $\Sigma 3$ grain boundaries in FCC metals

Jonathan Humberson[a] and Elizabeth A. Holm[a,*]

[a]Department of Materials Science and Engineering, Carnegie Mellon University, Pittsburgh, PA 15213 USA

**Abstract**

Synthetic driving force molecular dynamics simulations were utilized to survey grain boundary mobility in three classes of incoherent $\Sigma 3$ twin boundaries: <112>, <110>, and <111> tilt boundaries. These boundaries are faceted on low energy planes, and step flow boundary motion occurs by glide of the triplets of partial dislocations that comprise the mobile facets. Systematic trends with inclination angle are identified and characterized. Observations of thermally activated, anti-thermal, and athermal motion are explained in terms of the orientation of the Shockley partial dislocations along close-packed and non-close-packed directions. Thermally activated boundaries follow a compensation effect associated with a facet roughening transition. As for all faceting boundaries, system size and driving force must be chosen with care to prevent simulation artifacts.

---

* Corresponding author. Tel: +1 412 268 1762; email: eaholm@andrew.cmu.edu

# 1 Introduction

The presence of Σ3 grain boundaries (GBs) has a profound effect on the properties of FCC materials [1-4], and the mobility of these boundaries plays an important role in microstructural evolution [5]. A molecular dynamics (MD) survey of GB mobilities in nickel by Olmsted et al. [6] revealed surprising behavior in a number of Σ3 GBs with boundary plane inclinations different from the coherent twin. Not only did these GBs show some of the highest mobilities of all the GBs studied, but a large number of them exhibited an anti-thermal variation of mobility with temperature, in which GB mobility decreases with increasing temperature, in contrast to typical models of thermally-activated GB motion [7-9]. The Olmsted survey identified these boundaries, but determining the cause of this behavior was outside the scope of their survey. Further questions were raised by Homer et al. [10, 11], who noted that the Σ3 GBs that exhibited thermally-activated behavior all possess a <110> tilt axis, whereas nearly all the other Σ3 boundaries moved anti-thermally. More recently, we [12] explored the motion of a single anti-thermal boundary and determined that the boundary facets along {111} coherent twin planes and more mobile {110} planes, and that motion of the boundary occurs by glide of the triplets of Shockley partial dislocations that make up this {110} facet. Additionally, we identified a secondary faceting transition in the mobile facet that limits its mobility at lower temperatures. In this paper, we extend these simulations to address a broader range of Σ3 GBs, to examine the mechanisms of their motion, and to compare several subsets of Σ3 GBs, including the curious difference in thermal behavior.

# 2 Models and methods

## 2.1 *Crystallography and structure of the studied boundaries*

The Σ3 boundaries studied here belong to three crystallographically related groups. If the coherent twin boundary (CTB) with {111} boundary plane orientation is rotated about the <110> direction that lies in the boundary plane, it creates a series of grain boundaries that share a common <110> tilt axis, until a 90° rotation reaches a {112} orientation, which is one of the symmetric incoherent twin boundaries (SITB). Similarly, if the coherent twin is rotated about the <112> direction that lies in the boundary plane, it creates a series of boundaries with a common <112> tilt axis, until a 90° rotation reaches a {110} orientation, another SITB. Lastly, if a SITB with a {112} boundary plane is rotated about the <111> direction that lies in the plane, it creates a series of boundaries that share a common <111> tilt axis, until a 90° rotation reaches the {110} SITB. These rotations, along with the relationship between these planes, are illustrated in figure 1. The boundaries with <110>, <112> and <111> tilt axes produced by these rotations correspond to boundaries lying on the edges of the grain boundary plane orientation fundamental zone proposed by Homer et al. [11] In this work, we will principally be concerned with those boundaries lying along the <110> and <112> tilt axes, though we will include one example of a boundary along the <111> tilt axis.

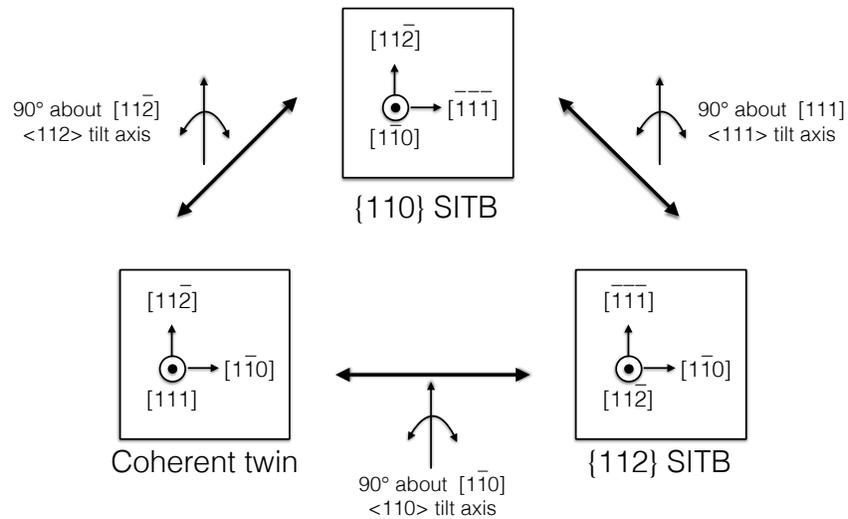

Figure 1. Schematic illustration of the crystallographic relationships between the Σ3 grain boundaries examined in this paper.

Researchers such as Banadaki and Patala [13] and Wang et al. [14, 15] have shown that the {111}, {112} and {110} boundaries represent local energy minima in the Σ3 boundary plane space (if there is no grain boundary dissociation such as that which occurs in low stacking fault energy materials like Cu) and that the structure and energy of general Σ3 boundaries are well-represented by a model in which the boundaries facet along those low-index planes. For boundaries constructed and minimized as described previously [12], we observe the predicted faceting behavior: <112> tilt boundaries facet on {111} and {110} planes; <110> tilt boundaries facet on {111} and {112} planes; and <111> tilt boundaries facet along {110} and {112} planes, as shown in figure 2. In each case, boundary motion is found to occur by step flow of the more mobile facet, as discussed below. It is worth noting that boundaries close to the CTB (<112> and <110> tilt boundaries with inclinations less than about 40° from the CTB) completely facet along the coherent twin plane, while boundaries with larger inclinations relative to the coherent twin instead form a series of atomic-scale twin facets along their length [16]. Nonetheless, when we create these high inclination boundaries in fully faceted forms, their mobilities are the same as for the unfaceted structures. This reflects that the motion of the boundary is controlled by the glide of Shockley partial dislocation triplets [14], whether they have aggregated into a single facet or remain distributed throughout the boundary.

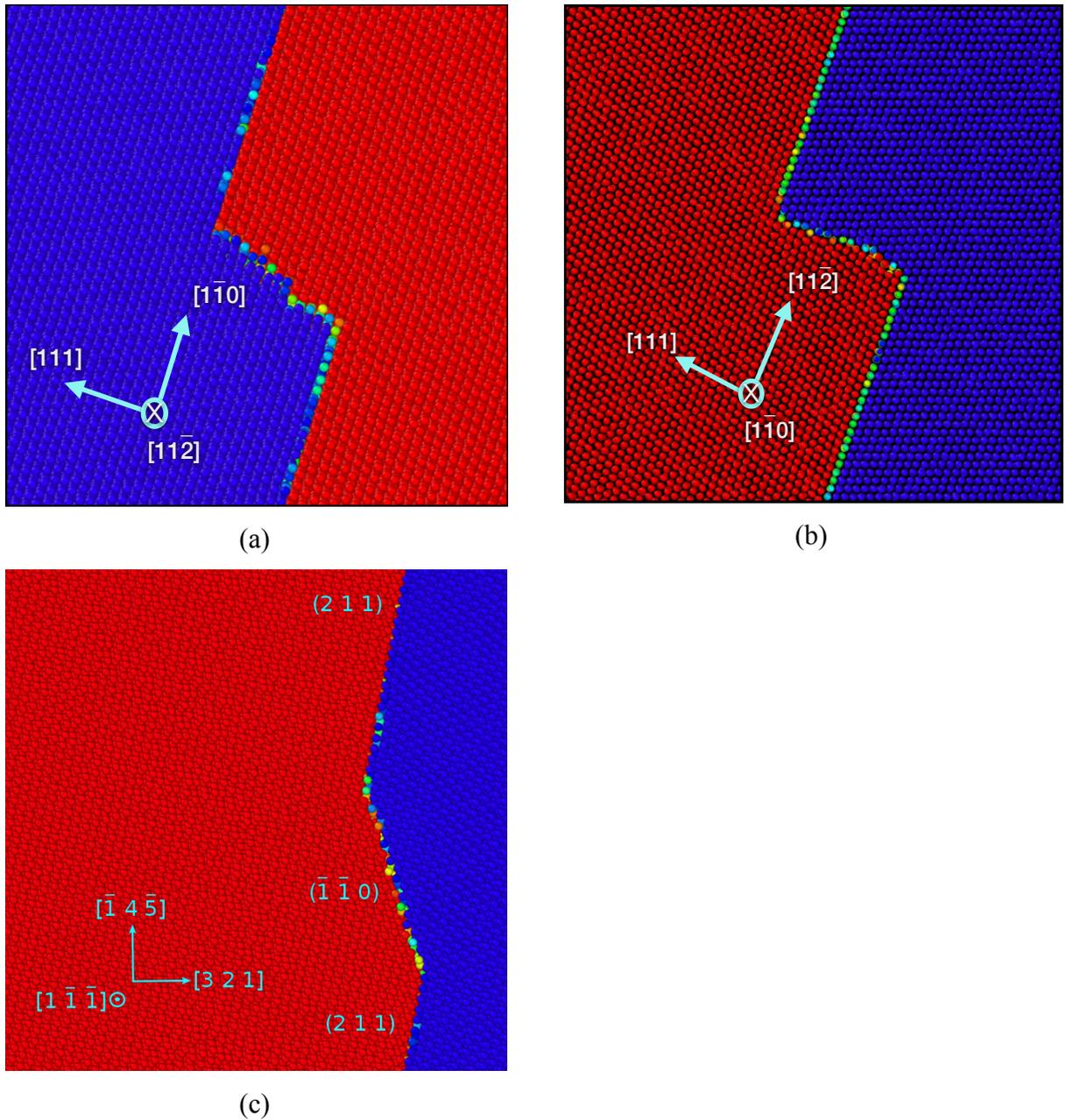

Figure 2. Faceting of incoherent Σ3 grain boundaries. (a) <112> 17.0° tilt boundary with {11 8 5} boundary normals, equilibrated at 1000K and viewed in a <112> direction. (b) <110> 25.2° tilt boundary with {10 4 4}/{8 8 2} boundary normals, equilibrated at 1000K and viewed in a <110> direction. (c) <111> tilt boundary with {3 2 1} boundary normals, equilibrated at 600K and viewed in a <111> direction. {112} and {110} SITB facets are indicated. Atoms are colored by the ECO order parameter, where red atoms represent the reference orientation.

## 2.2 Parameters for boundary simulation

In these simulations, we employed the same Foiles-Hoyt embedded atom method (EAM) interatomic potential for nickel [17] that was used in the earlier grain boundary mobility survey [6]. The simulation cell is constructed using GBpy, a Python package for calculating the geometric properties of bicrystals [18]. It is periodic in all three Cartesian directions, with a length normal to the boundaries of 218.6Å; the dimensions in the plane of the boundary and the magnitude of the driving force vary as described below. The ECO driving force [19] was used to drive the motion of the grain boundaries in the LAMMPS molecular dynamics code package [20, 21]. A cutoff radius of 1.1 lattice parameters (3.872Å) was used, sufficient to include both first and second nearest neighbors. This cutoff was found to properly distinguish between atoms in each grain up to the maximum temperature of 1400K, ensuring that the nominal applied driving force was representative of the actual driving force on the boundary. A order parameter cutoff value of $\eta = 0.25$ was used. The systems were maintained at their target temperature by a Nosé-Hoover thermostat [22, 23] and at zero pressure by a Parrinello-Rahman barostat with the modifications of Martyna, Tobias, and Klein [24]. Mobility was calculated from boundary displacement versus time data via bootstrap resampling [25] with a smoothing window of 5 ps and a sample window of 20 ps. We note that bootstrap resampling provides an accurate estimation of mobility for a relatively small data set, at the cost of underestimating the standard deviation [25].

As we and others have emphasized [12, 26, 27], the appropriate choice of simulation parameters, specifically the simulation dimensions in the grain boundary plane and the magnitude of driving force used, is crucial for physically accurate simulation of boundary motion. For the simulation length in the principal faceting direction, i.e. the dimension that determines the maximum size of the CTB facet, we must consider both the interaction between the facet junctions and the number of dislocation triplets that make up the mobile facet. We previously determined that the number of dislocation triplets seems to be the more important of these two factors, and so in this work we choose a simulation size in this direction sufficient to yield at least five repeats of the grain boundary period.

We also showed that the simulation length along the tilt axis of <112> tilt boundaries determines the maximum possible facet length when the mobile {110} section facets into {112} sections, and that this affects the temperature at which the faceting transition occurs. However, given the apparent difficulty of observing the {110} to {112} faceting transition in our system [12], we choose a simulation size along the <112> direction to give at least six repeats of the grain boundary period, found to be free of size effects above the faceting transition temperature. We then perform synthetic driving force molecular dynamics simulations to determine the grain boundary mobility down to 400K, with the knowledge that at some temperature below 600K the {110} sections will facet along {112} planes, and the mobility will drop abruptly. We use the same conditions to determine the size for the <110> tilt boundaries, though it should be noted that no faceting transition will occur for these boundaries, because the mobile facet is already the lower energy {112} boundary.

To determine appropriate synthetic driving forces, we note that all of the boundaries sharing a <112> tilt axis move via the glide of dislocation triplets that make up the {110} facet, as in the boundary studied previously [12]. Because these boundaries all share a common motion mechanism, we choose the driving force found to yield a mobility consistent with that given by zero-driving-force fluctuation methods, 1 meV/atom (14.7 MPa), for all the <112> tilt boundaries, which is considerably smaller than in typical MD studies [6, 28, 29]. For the <110> tilt boundaries, however, the dislocation triplets that make up the {112} facet have a much lower mobility than those in the {110} facet, and so require a correspondingly higher driving force. To determine an appropriate driving force for these boundaries, we took a representative <110> tilt boundary with {5 5 2}/{211} boundary normals and simulated the motion of the boundary at 700K using a range of driving forces, the results of which are shown in figure 3. At high driving forces, there is a strong, systematic variation in mobility with driving force, similar to that seen in other MD studies of grain boundary motion [6, 19, 29]. At low driving forces, the {5 5 2}/{211} boundary moves so little over the course of the simulation that the calculated mobility is not significantly different from zero. In the range of 5 meV/atom to 10 meV/atom, however, the mobility does not change with driving force, and so we choose 10 meV/atom (147 MPa) as the driving force for the <110> tilt boundaries in the interest of moving the boundary as far as possible during the simulation.

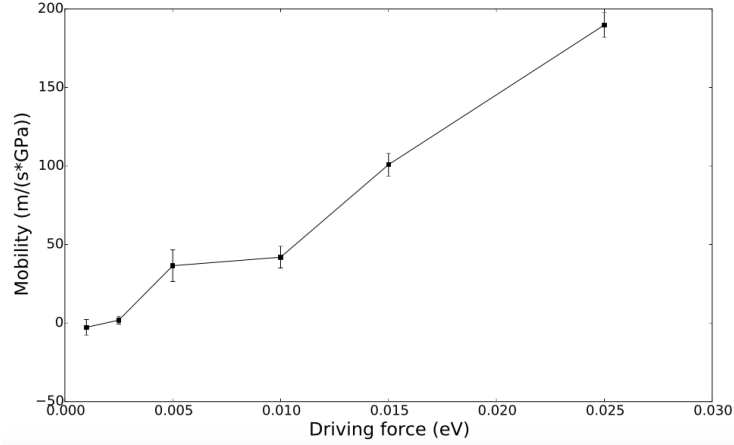

Figure 3. Variation of mobility with driving force for a <110> 19.7° tilt boundary with {5 5 2}/{211} boundary normals at 700 K.

## 3 Results and Discussion

### 3.1 Thermal behavior of <112> tilt boundaries

With the appropriate parameters in place, we simulate the mobility of a series of <112> tilt boundaries as given in table 1. For each boundary, motion is simulated and mobility is calculated at temperatures from 400K to 1400K at intervals of 100K, as shown in figure 4(a).

Table 1: Crystallographic details of simulated Σ3 <112> tilt boundaries.

| Boundary planes | Inclination angle relative to coherent twin | Boundary number in Olmsted survey [6] |
| --- | --- | --- |
| {11 8 5}/{11 8 5} | 17.0° | 366 |
| {7 4 1}/{7 4 1} | 31.5° | 45 |
| {5 2 1}/{5 2 1} | 50.7° | 11 |
| {6 2 2}/{6 2 2} | 58.5° | 47 |
| {8 4 2}/{8 4 2} | 67.8° | 78 |
| {11 7 2}/{11 7 2} | 74.8° | 258 |
| {1 1 0}/{1 1 0} | 90.0° | 5 |

The most immediately striking feature of these results is the similarity in the variation of mobility with temperature across all the boundaries, a reflection of the dislocation triplet motion mechanism common to these <112> tilt boundaries. In each case, we see a strongly anti-thermal trend in the mobility, with mobilities for many boundaries reaching several thousand m/(s·GPa) at low temperature. (It should be reiterated that we expect the mobile {110} facet to undergo a structural transition to {112} facets at some temperature below 600K, at which point the mobility of the boundary is expected to drop abruptly [12, 30, 31].)

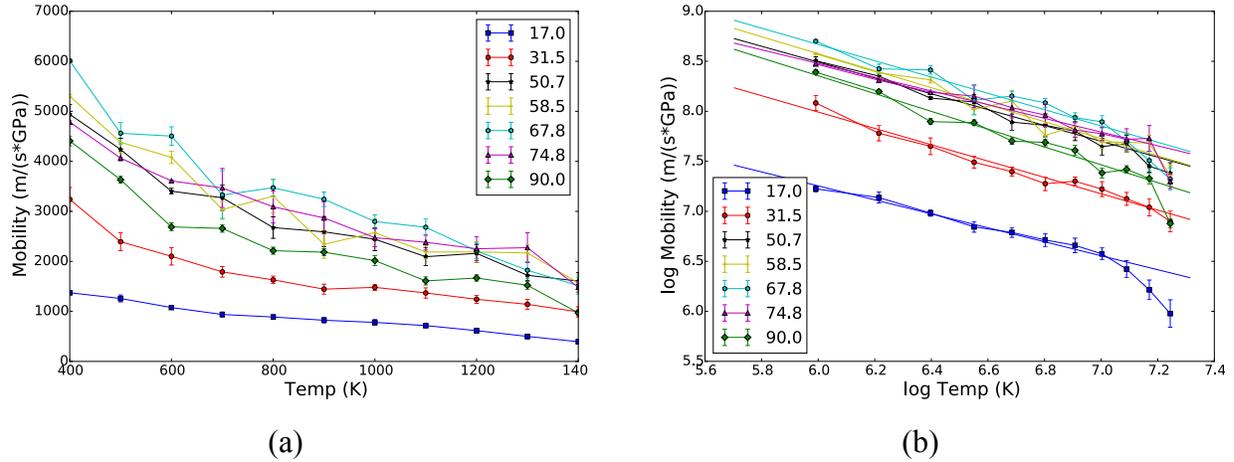

(a)                                         (b)

Figure 4: The variation of mobility with temperature for several Σ3 <112> tilt boundaries (a) plotted on linear axes and (b) in logarithmic coordinates, with linear fit is for temperatures ≤ 1200K. Insets give the angle in degrees that each boundary makes with the coherent twin.

In figure 4(b), we plot the logarithms of temperature and mobility. To avoid the drop in mobility seen at temperatures nearing the melting point $T_m$ = 1565K for Foiles-Hoyt Ni [17], linear regression was performed only on points at 1200K and below. In each case, mobility was found to be a power law in temperature up to about $0.8T_m$. Referring to the slopes given in table 2, we see that the power law exponents are in the range from -0.7 to -0.9. Although this differs from the theoretical value of -1 for glide of a lone dislocation, there are additional factors affecting the rate of motion of a grain boundary: The ability of an individual dislocation in the boundary to advance is not only determined by the lattice through which it moves, but also by the requirement that the dislocations in a given triplet move together, and beyond that by the relative position of neighboring dislocation triplets. In light of this, it is unsurprising that we see results that look qualitatively similar to, but are quantitatively different from, the results expected for a single dislocation.

Table 2: Power law curve fits for mobility versus temperature in figure 4(b).

| Inclination angle relative to coherent twin | Slope of best-fit line | Correlation coefficient, $r^2$ |
|---|---|---|
| 17.0° | -0.699 | 0.981 |
| 31.5° | -0.817 | 0.971 |
| 50.7° | -0.792 | 0.981 |
| 58.5° | -0.849 | 0.942 |
| 67.8° | -0.814 | 0.952 |
| 74.8° | -0.686 | 0.986 |
| 90.0° | -0.900 | 0.968 |

### 3.2 Angular variation of <112> tilt boundary mobility

In addition to considering how the mobility varies with temperature across a series of boundary inclinations, we may also consider the complementary question of how mobility varies with the inclination of those boundaries at a number of different temperatures. In each of these <112> tilt boundaries, we recall that the boundary motion occurs by motion of the {110} facets, while the CTB facets remain immobile. Thus, a natural expectation would be that as the angle a boundary makes with the coherent twin increases, the length of CTB facets decreases and the length of {110} facets increases, and this should produce a corresponding increase in the boundary mobility. We plot the data in this fashion in figure 5, and observe that this trend generally holds true at lower angles, but in every case the {110}/{110} SITB (inclined at 90° to the coherent twin) has a mobility considerably lower than the trend would suggest. This may be understood by considering the difference in structure between the Σ3 {110}/{110} boundary and the rest of the <112> tilt boundaries. For the Σ3 {110}/{110} boundary to move forward from its initially planar state, one of the dislocation triplets from which it is formed must first advance alone. For this triplet to move ahead of the rest of the boundary, however, requires the formation of new sections of CTB, which raises the energy barrier to this first step of forward motion. In contrast, a boundary with an amount of geometrically necessary CTB content always has a junction between the CTB and {110} facets (specifically, the junction that is concave with respect to the direction of boundary motion), at which the advance of a dislocation triplet does not require the creation of any CTB, and therefore that boundary has a much lower energy barrier to forward motion and a higher mobility.

The idea that the mobility of a <112> tilt boundary is tied to the relative amounts of CTB and {110} facet that comprise the boundary motivates us to ask if a geometric model such as that used by Tschopp and McDowell [32-35] to describe Σ3 GB energies may describe the variation in mobility. We attempted to fit a model analogous to that used by Tschopp and McDowell to our data with the form $M(\theta) = M_0 \sin(\theta)$, but found the fit to be generally poor, as it did not capture either the linear increase in mobility near the CTB nor the decrease in mobility approaching the {110} SITB. We conclude that a simple geometric model of the kind that proved successful in describing the energies of Σ3 GBs is insufficient to describe their mobilities; any model attempting to do so must at least incorporate a term accounting for the decrease in mobility as the boundary inclination approaches that of the {110} SITB.

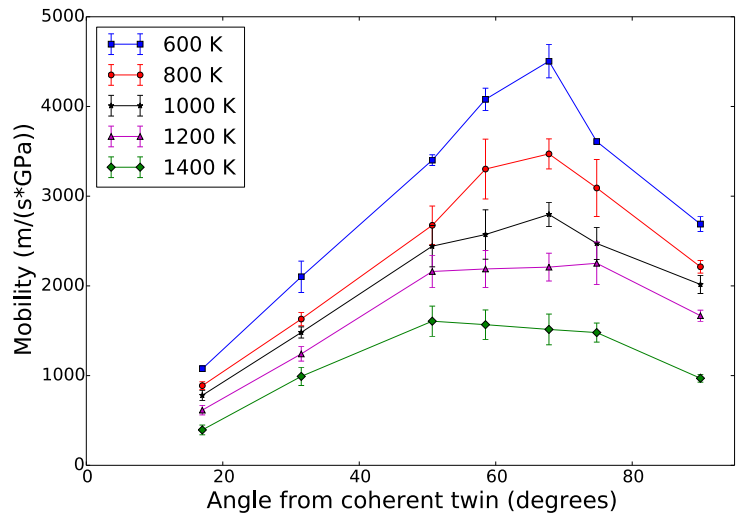

Figure 5: The variation of mobility of Σ3 <110> tilt boundaries with inclination to the coherent twin. The inset gives the temperature for each series.

### 3.3 Thermal behavior of <110> tilt boundaries

We now turn our attention to the other subset of Σ3 boundaries principally of interest: the <110> tilt boundaries. The structures and energies of Σ3 <110> tilt boundaries have been well-investigated, particularly the {112} SITB, which plays a large role in twinning and detwinning processes [15, 36-40]. Here, as we did for the Σ3 <112> tilt boundaries, we simulate the motion of <110> tilt boundaries at a number of inclinations to the coherent twin over a range of temperatures, as given in table 3. Owing to the lower mobility of these <110> tilt boundaries relative to the <112> tilts, we will concern ourselves with their thermal behavior in the range of 800K - 1400K. Even at these relatively high temperatures, several boundaries show so little motion that we cannot be confident that the calculated mobilities are accurate, and so we disregard any mobility found to be below 20 m/(s·GPa). Notably, the {112} SITB, at an inclination of 90° to the coherent twin, was found to be immobile at temperatures below 1200K, and so we simulate its motion at increments of 50K, rather than the increment of 100K used throughout the rest of this work. The results of these simulations are shown in figure 6(a).

Table 3: Crystallographic details of simulated Σ3 <110> tilt boundaries.

| Boundary planes | Inclination angle relative to coherent twin | Boundary number in Olmsted survey [6] |
|---|---|---|
| {5 5 2}/{2 1 1} | 19.7° | 119 |
| {10 4 4}/{8 8 2} | 25.2° | 159 |
| {4 1 1}/{1 1 0} | 35.2° | 20 |
| {14 2 2}/{10 10 2} | 43.3° | 333 |
| {8 1 1}/{5 5 4} | 64.6° | 163 |
| {1 1 2}/{1 1 2} | 90.0° | 4 |

The qualitative difference between the thermal behavior of these <110> tilt boundaries and that of the <112> tilts is immediately apparent. Whereas the <112> tilt boundaries uniformly showed a strong anti-thermal trend and exceptionally large mobilities, these <110> tilts show a thermally-activated trend and much smaller mobilities. To determine the activation energies for boundary motion, we replot mobility in Arrhenius coordinates in figure 6(b) and summarize the best linear fits in table 4.

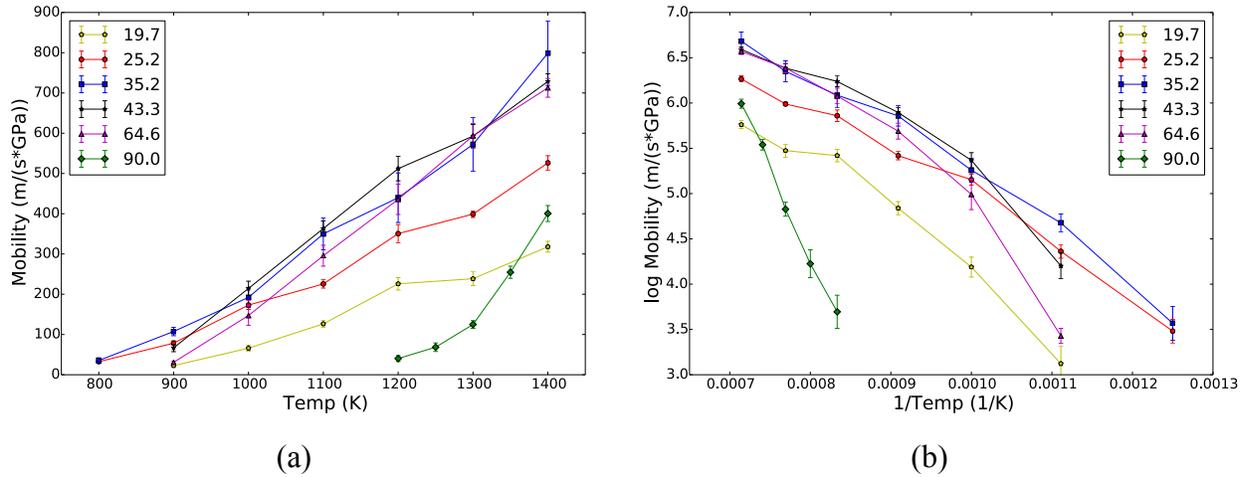

Figure 6: The variation of mobility with temperature for several Σ3 <110> tilt boundaries (a) plotted on linear axes and (b) in Arrhenius coordinates. Insets give the angle in degrees that each boundary makes with the coherent twin.

While the {112} SITB shows a linear dependence of log mobility on inverse temperature across the entire temperature range, figure 6(b) shows that the other <110> tilt boundaries have two distinct linear regions with different slopes. This behavior has been observed previously, where the change in activation energy is presumed to correspond to a change in atomic motion mechanism [29]. Here, excluding the {112} SITB, we find that the activation energies at high temperatures (averaging 0.33 eV/atom) are smaller than those at low temperatures (averaging 0.71 eV/atom). The change in activation energy is found to occur between 1000K and 1200K, and the transition temperature does not vary in a systematic manner with respect to inclination angle (or, equivalently, {112} SITB boundary content). Both the decrease in activation energy as temperature increases and the magnitudes of the transition temperatures are consistent with a thermal roughening transition [29]. While the CTB facets remain definitively flat at all temperatures, roughening of the {112} planes is plausible and could take the form of increased separation within or between the Shockley partial dislocation triplets that comprise the {112} facets. Because of the small system sizes in these simulations, such a transition would be difficult to measure directly [29], but the activation energy change provides circumstantial evidence.

The {112} SITB stands out from the rest of the <110> tilt boundaries in that its activation energy for motion is considerably larger than that of any of the other boundaries. However, we note that the {112} SITB was found to be immobile at MD time scales below 1200K, suggesting that this boundary may undergo a similar transition, but that the resulting mobility is so low that simulating the motion of the boundary is outside of the time scale accessible by MD.

The activation energies and prefactors in table 4 allow us to check for a compensation effect [7, 41, 42], i.e. whether these two quantities are linearly related. As plotted in figure 7, with the exception of the {112} SITB, the data from both the high and low temperature regimes fall along the line $E_a = 0.0915 M_0 - 0.49759$. This gives a compensation temperature, at which all processes occur at the same rate, of $T_c = 1063$K, which agrees well with the observed roughening transition temperatures.

Table 4: Mobility prefactor and activation energy calculated from linear fits to Arrhenius plots of $\Sigma 3$ <110> tilt boundary mobilities.

| Boundary planes | Inclination angle relative to the CTB | Temperature range (K) | Logarithmic prefactor $M_0$ (log m/(s·GPa)) | Activation energy $E_a$ (eV) |
|---|---|---|---|---|
| {5 5 2}/{2 1 1} | 19.7° | ≤ 1200 | 12.29 | 0.707 |
| {5 5 2}/{2 1 1} | 19.7° | ≥ 1200 | 7.73 | 0.243 |
| {10 4 4}/{8 8 2} | 25.2° | ≤ 1000 | 11.81 | 0.575 |
| {10 4 4}/{8 8 2} | 25.2° | ≥ 1000 | 9.05 | 0.338 |
| {4 1 1}/{1 1 0} | 35.2° | ≤ 1100 | 11.89 | 0.569 |
| {4 1 1}/{1 1 0} | 35.2° | ≥ 1100 | 9.61 | 0.360 |
| {14 2 2}/{10 10 2} | 43.3° | ≤ 1100 | 13.67 | 0.729 |
| {14 2 2}/{10 10 2} | 43.3° | ≥ 1100 | 9.07 | 0.298 |
| {8 1 1}/{5 5 4} | 64.6° | ≤ 1100 | 16.07 | 0.973 |
| {8 1 1}/{5 5 4} | 64.6° | ≥ 1100 | 9.87 | 0.394 |
| {1 1 2}/{1 1 2} | 90.0° | ≥ 1200 | 20.33 | 1.728 |

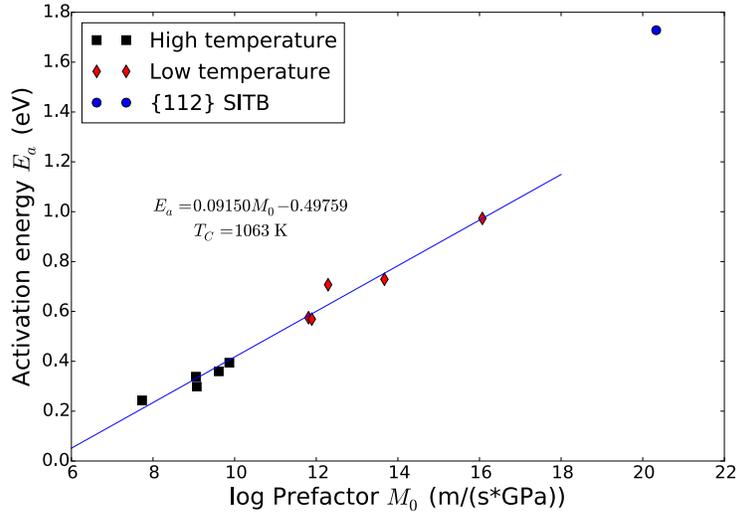

Figure 7. Compensation effect between activation energy $E_a$ and logarithm of mobility prefactor $M_0$ for <110> tilt boundaries. The compensation temperature $T_c$ = 1063K.

### 3.4  Angular variation of <110> tilt boundary mobility

The dependence of mobility on inclination angle with respect to the coherent twin is presented in figure 8. We observe a trend qualitatively similar to the one seen in figure 5 for <112> tilt boundaries, where the boundary mobility initially increases with inclination as would be predicted by a geometric model, but drops off quickly as the inclination approaches the SITB at 90°. Here, the drop in mobility for the {112} SITB is larger than the corresponding drop for the {110} SITB as a result of the lower mobility of the {112} SITB. We again conclude that a simple geometric model is insufficient to capture the variation of mobility with boundary inclination for <110> tilt boundaries.

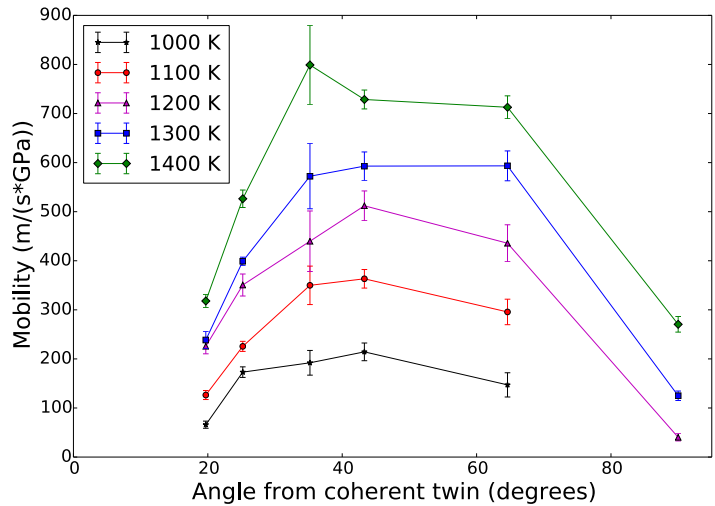

Figure 8: The variation of mobility of Σ3 <110> tilt boundaries with inclination to the coherent twin. The inset gives the temperature for each series.

### 3.5 *Differences in thermal behavior between <110> tilt and <112> tilt boundaries*

The <112> and <110> tilt boundaries have much in common structurally. Both facet strongly along the CTB and a SITB, and motion of the boundary as a whole occurs by motion of the much more mobile SITB facet. Both the {110} SITB and the {112} SITB consist of the same triplets of Shockley partial dislocations. There is, however, a striking difference not only in the magnitudes of their mobilities, but also in the variation of those mobilities with temperature.

We typically regard screw and mixed dislocations to be less mobile than edge dislocations. However, this understanding comes from experiments on and simulations of perfect dislocations, albeit perfect dislocations that have dissociated into Shockley partial dislocations, as is typical in FCC materials. In FCC materials, a perfect dislocation has a Burgers vector of the type $a_0/6<110>$; as a consequence, screw and 60° mixed dislocations with these Burgers vectors must point along <110> directions, which are the most closely-packed directions in the {111} planes. Correspondingly, edge and 30° mixed dislocations must point along <112> directions, which are the second most closely-packed directions in these planes. This prompts us to ask if there is an effect on dislocation mobility, separate from the effect of dislocation character, that arises from the orientation of a dislocation's line vector in the {111} plane. Such an effect would presumably arise from the difference in the Peierls stress and energy of dislocations with different orientations, as a result of the different atomic packing along different directions. There are some studies in the literature that allow us to explore the effect of dislocation orientation, and we summarize three relevant results here.

First, Schoeck and Krystian [43] used numerical calculations in the Peierls-Nabarro model to determine the Peierls energies for dissociated screw, 30°, 60°, and edge dislocations in Cu, and found that the screw and 60° dislocations both have Peierls energies much higher than those of the edge and 30° dislocations. Second, Lu et al. [44] used both density functional theory (DFT) and EAM calculations to construct generalized stacking fault (GSF) surfaces, as proposed by Vitek and Cockayne [45, 46]. These GSF surfaces were then used in a modified version of the Peierls-Nabarro model, developed by the authors, to determine Peierls stresses for the screw, 30°, 60°, and edge dislocations. Similarly to Schoeck and Krystian, the authors found the Peierls stresses of the screw and 60° dislocations to be considerably higher than those of the edge and 30° dislocations. Last, and most strikingly, Szelestey et al. [47] performed a series of MD simulations in which a dislocation of either edge or screw character, dissociated into its component partial dislocations, was placed in a system with fixed simulation boundaries in the direction of dislocation glide. By tracking the peaks in the atomic misfit function, the authors were able to track the positions of each partial dislocation independently, and by applying a stress to the simulation cell and accounting for the image force from the fixed boundaries, they were able to track the displacement of each partial dislocation as a function of applied stress and find the Peierls stress for each partial dislocation independently. It should be emphasized that the partial dislocations produced from the dissociation of the edge and screw dislocations have the same Burgers vectors; the only difference is the orientation of the dislocations. In keeping with the results of the other two studies, when the partial dislocations are oriented along the close-packed <110> direction, as in the screw case, the resulting Peierls stress is over an order of magnitude higher than when they are oriented along the <112> direction, as in the edge case. Additionally, the difference in Peierls stress from partial dislocation orientation is higher than might be inferred from the effective Peierls stresses of the edge and screw dislocations. This is a result of the separation of partials for the screw dislocation not being an integer multiple of the atomic spacing, leading to a partial cancellation of the Peierls stresses of the partials as one partial "helps" the other over the Peierls barrier, which produces a lower effective Peierls stress for the dissociated dislocation. In the edge dislocation, the separation of partials is approximately an integer multiple of the atomic spacing, and so the two partials move in phase and no cancellation occurs.

In summary, studies that address the effects of partial dislocation line vector show a large effect on the Peierls stress or energy stemming from the orientation of dislocations, independent of their character. This is not to say that dislocation character has no effect on these properties, of course; the works of Schoeck and Krystian [43] and of Lu et al. [44] both show differences between screw and 60° dislocations and between edge and 30° dislocations, and these differences presumably stem from the characters of these dislocations. Nonetheless, the results of these studies unanimously indicate that dislocations (and partial dislocations) oriented along a <110> direction have a considerably higher barrier to motion than those oriented along a <112> direction. If this is the case, this effect would help to explain the observed difference in boundary mobility, since the partial dislocations in the {110} SITB all point in the second most closely-packed <112> directions, whereas the partial dislocations in the {112} SITB all point in the most closely-packed <110> directions.

### 3.6 A <111> tilt grain boundary

In order to elucidate the effects of the CTB facets on boundary motion, we consider a $\Sigma 3$ <111> tilt boundary with {3 2 1} boundary normals (number 30 in the Olmsted survey [6]). In agreement with the results of Banadaki and Patala [13], this boundary facets along the {112} and {110} SITBs, as seen in figure 2(c). Similarly to the <110> and <112> tilt boundaries, one facet (the {110} SITB) is much more mobile than the other, and motion of the boundary proceeds via the motion of this mobile facet. Because the boundary is comprised of facets that contain the same dislocations with different orientations, the motion of this boundary is similar to the motion of persistent dislocation kinks.

Since boundary motion resembles dislocation kink propagation, we expect the energy barrier to boundary motion will be small, and so we must determine the effect of the magnitude of driving force on the boundary mobility. The calculated mobilities for a range of driving forces are shown in figure 9(a). At high driving forces, the boundary is overdriven, resulting in a mobility that is lower than observed for smaller driving forces. Below a driving force of 1 meV/atom, the boundary moves so little that the error estimates dwarf the mobilities. Based on this, we choose a driving force of 1 meV/atom for this boundary.

We must also consider the size of the simulation in the [-1 4 -5] direction, the direction along which the boundary facets into {112} and {110} SITBs. Rather than determining the number of dislocation triplet units in the mobile facet, as was the case for the <110> and <112> tilt boundaries, here the size of the simulation in the faceting direction determines the length of the mobile dislocation kink. The variation of calculated mobility with system size in this direction is shown in figure 9(b). A modest effect of size on mobility is evident at small system sizes. Thus, we choose a length of 10 GB periods, or approximately 114Å, in the [-1 4 -5] direction. Along the tilt axis, we choose a length of 6 periods, or 18.4Å. It should be noted that the length along the tilt axis does not influence the faceting behavior of the boundary, as it does for the <112> tilt boundaries. (Note, however, that we do expect a low temperature faceting transition in this boundary, with the {110} SITB facet breaking up into {112} SITB facets, though the difference in boundary structure may cause this transition to occur at a different temperature than in the <112> tilt boundaries.)

Using these simulation parameters, we determine the mobility of the <111> tilt boundary over the same temperature range as was used for the <110> and <112> tilt boundaries. As shown in figure 9(c). The boundary shows essentially no change in mobility with temperature over a range of 1100K, though the mobility drops by about 200 m/(s·GPa) at 1400K. Athermal boundary mobility is observed when the local driving force for GB motion is large enough that it overwhelms the activation barrier. For example, Kopetsky et al. [48] observed athermal motion in a Zn bicrystal with Bi solute, which the authors attributed to the boundary breaking away from the solute. Given that the <111> tilt boundary motion is similar to kink propagation, we can expect a very small activation barrier, and so overdriving may be a problem here, as well. While our exploration of the effect of driving force on mobility in figure 9(a) did not suggest that 1 meV/atom would overdrive the boundary, it is nonetheless possible that reaching the regime in which the motion of the boundary is truly a biased thermal process would require even smaller driving forces. If this is the case, then zero-driving-force fluctuation-based methods would be the most appropriate to recover the true mobility [49], although these methods are also challenged by fully faceted boundaries.

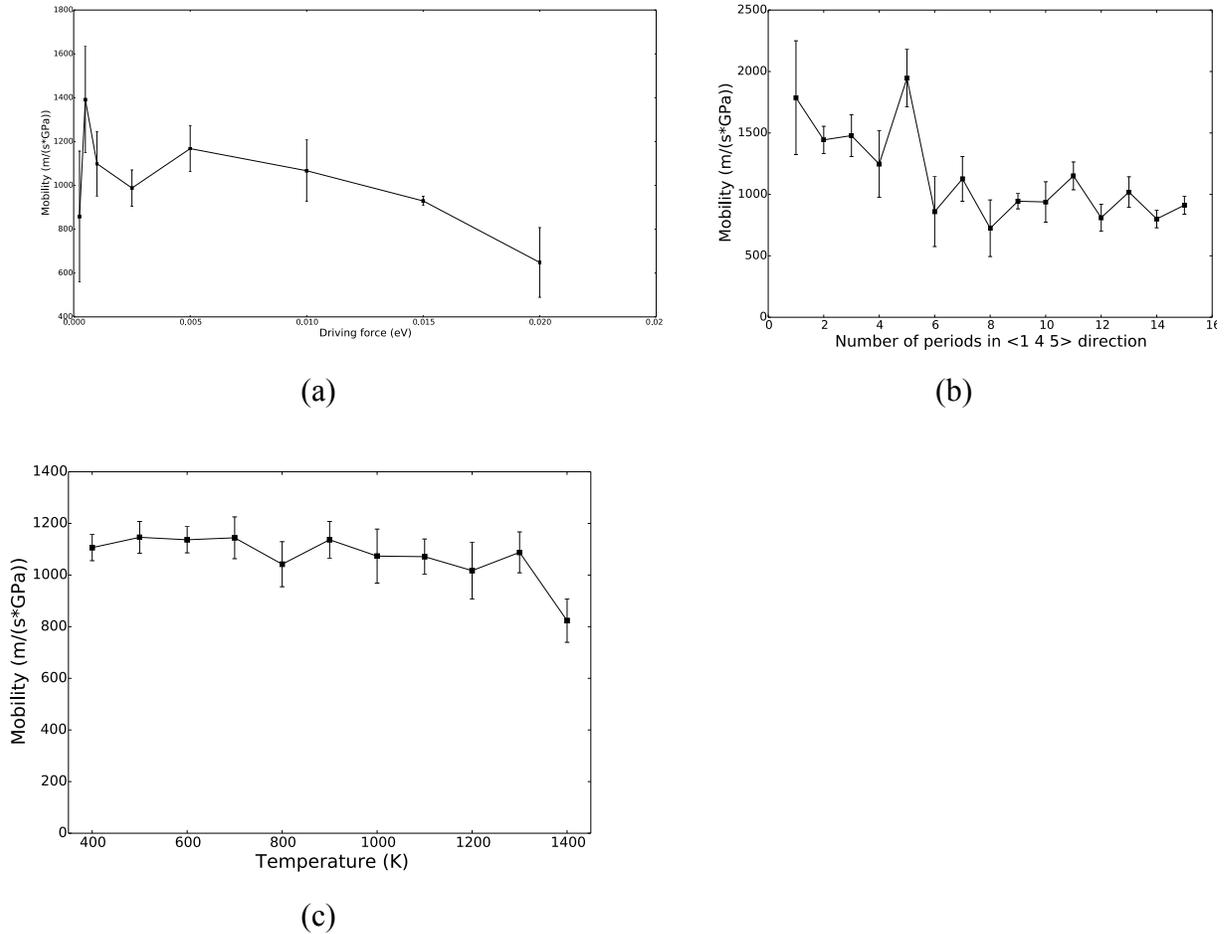

(a)

(b)

(c)

Figure 9: (a) Variation of calculated mobility of chosen <111> tilt boundary with (a) applied driving force at 700K, (b) number of grain boundary periods in the [-1 4 -5] direction at 700K, and (c) temperature.

## 4 Conclusions

Though the Σ3 boundaries show considerable anisotropy, this work reinforces the idea that their behavior may be understood in terms of a number of low-index, high-symmetry boundary planes, namely the {111}, {110} and {112} planes (c.f. [13]). In the cases of the <110> and <112> tilt boundaries, motion occurs by movement of the higher mobility {112} and {110} facets, respectively, while the less mobile {111} facet remains stationary. Similarly, in the case of the <111> tilt boundary studied, movement occurs by the motion of the more mobile {110} facet over the {112} facet. These results also echo the recent results of Hadian et al. [26] that in certain temperature regimes the flat Σ7 symmetric tilt boundary is immobile, but the perturbation of the boundary away from this low mobility inclination introduces geometrically necessary steps or kinks, which allow boundary motion with much lower activation barriers.

For the <110> and <112> tilt boundaries, mobility was found to vary smoothly with inclination angle to the coherent twin, initially increasing as the mobile facet normal gets closer to the direction of boundary motion before dropping near the SITB, which is a result of the consistency of motion mechanism both within each boundary set and between the two sets. This consistency is emphasized by the observation that boundaries inclined above about 40° with respect to the coherent twin do not display a persistent facet structure [16], but the presence or lack of persistent large facets does not affect the smooth variation in boundary mobility or the observed motion mechanism, i.e. the glide of triplets of Shockley partial dislocations. Additionally, we note that because the mobility of these boundaries varies smoothly with both temperature and angle to the coherent twin, a simple interpolation can provide mobilities as a function of temperature and inclination for input to mesoscale methods of simulation.

The striking difference in thermal behavior between the <110> and <112> boundaries, pointed out by Homer et al. [11], is surprising given their similarity in structure and motion mechanism. However, we understand this difference in terms of the orientation of the Shockley partial dislocations that comprise the mobile facets. In FCC materials, we do not often consider dislocation orientation to have such a large effect on the Peierls stress and energy, but the available information in the literature points to a consistently higher barrier for dislocations oriented in a close-packed direction. The collective effect of this increased barrier for each triplet of Shockley partials leads to a significant difference in these two sets of boundaries, with the <112> tilt boundaries undergoing anti-thermal motion with exceptionally large mobilities, and the <110> tilt boundaries evincing more typical thermally activated motion, with a thermal roughening transition at about 1100K.

Finally, in the <111> tilt boundary studied here, boundary motion occurs via dislocation kink propagation. The extremely low energy barrier for this process results in overdriven motion, and athermal mobility, even for the smallest driving forces accessible to synthetic driving force molecular dynamics simulations. The zero driving force properties of such boundaries remain a topic for future study.

## Acknowledgements

This work was supported by the National Science Foundation under Award Number DMR-1307138.